\documentclass[aps,pre,eprint,groupedaddress]{revtex4-2}

\usepackage{graphicx}
\usepackage{xcolor}
\usepackage{amsmath,amssymb,amsfonts}

\begin{document}

\title{$Q$-voter model with independence on signed random graphs:\\ approximate master equations}

\author{A.\ Krawiecki and T.\ Gradowski}       

\affiliation{Faculty of Physics,
Warsaw University of Technology, \\
Koszykowa 75, PL-00-662 Warsaw, Poland}

\begin{abstract}
Approximate master equations are derived for the two-state $q$-voter model with independence on signed random graphs, with negative and positive weights of links corresponding to antagonistic and reinforcing interactions, respectively. Depending on the mean degree of nodes, the size of the $q$-neighborhood, and the fraction of the antagonistic links, with decreasing independence of agents, this model shows a first- or second-order ferromagnetic-like transition to an ordered state with one dominant opinion. Predictions of the approximate master equations concerning this transition exhibit quantitative agreement with results of Monte Carlo simulations in the whole range of parameters of the model, even if predictions of the widely used pair and mean field approximations are inaccurate. Heterogeneous pair approximation derived from the approximate master equations yields results indistinguishable from homogeneous pair approximation studied before and fails in the case of the model on networks with a small and comparable mean degree of nodes and size of the $q$-neighborhood.
\end{abstract}

% multiplex networks; phase transitions; pair approximation; $q$-voter model.

\maketitle

% 111111111111111111111111111111111111111111111111

\section{Introduction}
\label{sec:intro}

An important class of models for social opinion formation \cite{Castellano09} consists of nonequilibrium agent-based models on networks, with agents exhibiting two possible, opposite opinions, represented by two-state spins, located in nodes and interacting via edges of a network (which can be a complete, regular, or random graph) imitating a, possibly complex, structure of social interactions \cite{Albert02, Dorogovtsev08}. Among them, one of the simplest is the $q$-voter (nonlinear voter) model with different forms of stochasticity \cite{Castellano09a, Nyczka12, Moretti13, Chmiel15, Jedrzejewski17, Peralta18, Jedrzejewski22, Gradowski20, Krawiecki24}. In this model, interactions are usually reinforcing, and the agents tend to change their opinions (spins tend to flip) if the opinion of a randomly selected subset of $q$ neighbors ($q$-neighborhood) is unanimously opposite. In the version of the $q$-voter model with independence, agents follow the above-mentioned rule with probability $1-p$ ($0\le p\le 1$) or act independently with probability $p$ and change or preserve their opinions randomly with equal probabilities $p/2$, where $p$ is a measure of internal noise and reflects agents' uncertainty in making decisions. The above-mentioned rules define the spin-flip rate for the $q$-voter model with independence. Related models for the opinion formation, e.g., the (linear) voter model \cite{Pugliese09, Carro16, Peralta18a, Baron21}, the majority vote model \cite{Oliveira92, Chen15, Krawiecki20, Kim21}, the $q$-neighbor Ising model \cite{Jedrzejewski15, Park17, Chmiel18, Krawiecki21, Krawiecki23}, etc., differ with spin-flip rates and the possible effect of internal noise. In general, these models predict the occurrence of a dominant opinion in the network of interacting agents with a decrease in internal noise level. For example, as $p$ is decreased, the $q$-voter model with independence exhibits phase transition from the paramagnetic (PM) state with both opinions equally probable to the ferromagnetic (FM) state with one dominant opinion, which can be first- or second-order, depending on the mean degree of nodes $\langle k \rangle$ in the network and the size of the $q$-neighborhood. 

Recently, the models mentioned above have been generalized to allow for antagonistic interactions associated randomly with a fraction $r$ of the links \cite{Krawiecki24, Baron21, Krawiecki20, Krawiecki21}. Then the underlying network can be treated as a signed network, with positive and negative links corresponding to reinforcing and antagonistic interactions, respectively. Generalization of the models is achieved by appropriate modifications of the spin-flip rates. For example, in the case of the $q$-voter model, the agent can change opinion if all members of the selected $q$-neighborhood have mismatched rather than opposite opinions, i.e., the same or opposite opinions if their interactions with the considered agent are antagonistic or reinforcing, respectively. Then, for the latter model on networks with fixed $\langle k \rangle$, $q$ phase diagrams on the plane $p$ vs.\ $r$ can be drawn, which reveal such phenomena as the appearance of a tricritical point (TCP) at a finite value of $r$ at which the critical lines for the first- and second-order FM transition merge, or disappearance of the FM transition with increasing $r$ \cite{Krawiecki24}. 

The process of opinion formation in the above-mentioned agent-based models has been studied extensively using Monte Carlo (MC) simulations and various theoretical approximations. In the case of models on complete graphs or complex networks with large mean degree of nodes $\langle k \rangle$ predictions of a simple mean-field approximation (MFA) usually reproduce quantitatively results of MC simulations, e.g., the order of the FM transition, the critical value(s) of the internal noise and the width of the possible hysteresis loop \cite{Castellano09, Nyczka12, Moretti13, Chmiel15, Chen15, Kim21, Jedrzejewski15}. In the case of models on complex networks with finite $\langle k \rangle$, predictions of the MFA can substantially differ from the results of MC simulations, and quantitatively better predictions are obtained using various forms of pair approximation (PA), mainly the simplest homogeneous PA \cite{Jedrzejewski17, Peralta18, Gradowski20, Jedrzejewski22, Krawiecki24, Carro16, Peralta18a, Krawiecki20, Chmiel18, Krawiecki21}. However, in some cases, predictions of the PA also become even qualitatively false. For example, in the case of the $q$-voter model with independence on signed complex networks with $r\ge 0$, and with small and comparable $\langle k\rangle$, $q$ the critical value(s) of $p$ and even the order of the FM transition are predicted incorrectly \cite{Gradowski20, Krawiecki24}. 

In this paper, a more detailed approach based on approximate master equations (AMEs) \cite{Gleeson11, Gleeson13, Peralta20, Unicomb18, Unicomb19} is applied to predict the critical behavior of the $q$-voter model with independence on signed networks in the range of parameters $\langle k\rangle$, $q$ where the homogeneous PA fails. The AMEs, originally formulated for models with binary dynamics on unweighted networks \cite{Gleeson11, Gleeson13}, are extended to take into account the presence of both reinforcing and antagonistic interactions, in analogy with AMEs for models on weighted networks \cite{Unicomb18, Unicomb19}. This approximation divides the nodes into classes depending on their degree and the number of attached negative (antagonistic) links. The dynamical variables are concentrations of spins oriented up or down, occupying nodes of each class, which have given numbers of neighboring spins oriented up and connected by edges of each kind. It is shown that, at the expense of a significant increase in the number of equations in comparison with the MFA and homogeneous PA, the AMEs quantitatively reproduce the phase diagrams for the FM transition on the $(p,r)$ plane in the whole range of the parameters of the model under study. The number of equations can be substantially decreased by introducing aggregate variables in the form of sums of the above-mentioned concentrations of spins over the numbers of neighboring spins oriented up, which leads to a sort of so-called heterogeneous PA \cite{Gleeson11, Gleeson13}. Unfortunately, it is shown that the latter approximation yields results indistinguishable from the previously considered signed homogeneous PA \cite{Krawiecki24}, a version of the homogeneous PA most suitable for models on signed networks, thus it is also incorrect in the case of the $q$-voter model with small and comparable parameters $\langle k\rangle$, $q$. This paper does not consider other, more accurate versions of the heterogeneous PA \cite{Pugliese09} and their possible extensions to the case of models on signed networks.

% 22222222222222222222222222222222222222222222222222222

\section{The model}

\label{sec:model}

The model considered in this paper is the $q$-voter model with independence \cite{Castellano09a, Nyczka12, Moretti13, Chmiel15, Jedrzejewski17, Peralta18, Jedrzejewski22} on signed random graphs, which is described in more detail in Ref. \cite{Krawiecki24}. In this model, agents are located in nodes of a random network, indexed by $j=1, 2, \ldots N$, with degrees $k_j$, and represented by spins $\sigma_j =\pm 1$ oriented up or down corresponding to two opposite opinions on a given subject. The network is characterized by the degree distribution $P(k)$ and the resulting mean degree of nodes $\langle k\rangle$. For reasons explained in Sec.\ \ref{sec:theory_AMEs}, in this paper, MC simulations and their comparison with theoretical results are performed only for the model under study on random regular graphs (RRGs) with $P(k)=\delta_{k, K}$, $\langle k \rangle = K$. The agents interact via links (edges) of the network. The interactions can be reinforcing or antagonistic, promoting parallel or antiparallel alignment of the interacting spins, respectively. The reinforcing and antagonistic interactions are associated with the links randomly, with probabilities $1-r$ and $r$, respectively. Then, the network of interactions can be formally viewed as a signed network, with a two-point distribution of weights $J_{j,j'} = \pm 1$ associated with the edges
\begin{equation}
P\left( J_{jj'} \right) = (1-r) \delta \left( J_{jj'}-1\right) + r \delta \left( J_{jj'}+1\right),
\end{equation}
thus the plus (minus) sign corresponds to the reinforcing (antagonistic) interaction. Hence, the number $l$ of links associated with antagonistic interactions attached to a node with degree $k$ obeys a binomial distribution $B_{k,l})(r) = {k\choose l} r^l (1-r)^{k-l}$. The opinions of the two interacting agents $\sigma_j$, $\sigma_{j'}$ are called mismatched if $J_{jj'}\sigma_{j} \sigma_{j'}=-1$, i.e., if they are opposite while the interaction between the agents is reinforcing or if they are identical while the interaction is antagonistic; otherwise, the opinions are called matched and there is $J_{jj'}\sigma_{j} \sigma_{j'}=+1$. 

The dynamics of the $q$-voter model with independence on signed random graphs generalizes that for this model on random graphs with only reinforcing interactions. During the MC simulation, the agents are picked randomly, and for each agent, a subset of $q$ its neighbors ($q$-neighborhood) is selected randomly. Then, the picked agent updates its opinion according to the following two rules, which depend on the parameter $p$ ($0 \le p \le 1$), measuring agents' independence in making decisions and, thus, the level of internal noise in the model. First, with probability $1-p$, the picked agent changes opinion (the corresponding spin flips) if the opinions of all members of the selected $q$-neighborhood are mismatched with its opinion. Second, with probability $p$, the picked agent changes opinion independently, regardless of the opinions of the members of the $q$-neighborhood. Thus, the flip rate for a spin occupying a node with degree $k$ and possessing $x$ neighbors with mismatched opinions is
\begin{equation}
    f(x|k) = (1-p) \frac{{x \choose q}}{{k \choose q}} + \frac{p}{2}.
    \label{rate_fxk}
\end{equation}

In the MC simulations, random sequential updating of the spins is performed according to the above-mentioned rules, and a single MC simulation step (MCSS) corresponds to the update of opinions of all $N$ agents without repetitions. Eventually, each MCSS is performed as follows.
\begin{itemize}
    \item[(i.)] A node $j$, $1\le j \le N$, with degree $k_j$ is picked randomly.
    \item[(ii.)] A set of its $q$ neighbors ($q$-neighborhood) is chosen randomly and without repetitions. It is assumed that $0<q\le k_j$; otherwise, the node is excluded from the simulation.  
    \item[(iii.)] With probability $1-p$, the picked agent flips its opinion if the opinions of all members of the chosen $q$-neighborhood are mismatched with its opinion. 
    \item[(iv.)] With probability $p$, the picked agent behaves independently and flips or preserves its opinion with equal probabilities $p/2$.
    \item[(v.)] Steps (i.)-(iv.) are repeated until all $N$ spins are updated without repetition.
\end{itemize}

% 3333333333333333333333333333333333333333333333333333

\section{Theory}

\label{sec:theory}

\subsection{Approximate master equations}

\label{sec:theory_AMEs}

The AMEs provide a detailed and accurate theoretical description of models with binary-state dynamics on networks, and both heterogeneous and homogeneous PA, as well as MFA, can be obtained from them as a series of consecutive approximations with decreasing accuracy. In the original formulation, two-state spins in such models are divided into classes according to their orientation, the degree of nodes they occupy, and the number of neighboring spins oriented up, and the AMEs have a form of master equations for the concentrations of spins in each class, with transition probabilities between different classes averaged over the entire network \cite{Gleeson11, Gleeson13}. The AMEs were then generalized to the case of models on weighted networks, with weight associated with links drawn from a finite set of values \cite{Unicomb18, Unicomb19}. Then, the set of nodes and the set of links associated with a given weight form a layer, and the union of all layers, for all possible weights, forms a multilayer network equivalent to the original weighted networks. The two-state spins are divided into classes according to their orientation, the degree of the node, and the number of neighboring spins oriented up within each layer, and the AMEs are again written as appropriate master equations for the densities of spins within each class. Following this idea, in this paper, the AMEs are derived for the $q$-voter model with independence on signed networks, which can be treated as weighted networks with two possible values of weights $\pm 1$.

Let $s_{k,l;m,n}(t)$ (resp.\ $c_{k,l;m,n}(t)$) denote the fraction of nodes with degree $k$ with $l$ attached negative links ($0\le l \le k$), occupied by spins oriented down (resp.\ up), which have $m$ neighboring nodes connected by positive links and occupied by spins oriented up ($0\le m\le k-l$) as well as $n$ neighboring nodes connected by negative links and occupied by spins oriented up ($0\le n\le l$). The AMEs for the above-mentioned concentrations have a form of master equations,
\begin{eqnarray}
\frac{d s_{k,l;m,n}}{dt} &=& - F_{k,l;m,n} s_{k,l;m,n}+ R_{k,l;m,n} c_{k,l;m,n} 
\nonumber\\
&-&  \beta_{s}^{(+)} \left(k-l-m\right)  s_{k,l;m,n}
- \beta_{s}^{(-)} \left(l-n\right)  s_{k,l;m,n} 
\nonumber\\
&+& \beta_{s}^{(+)} \left(k-l-m+1\right)  s_{k,l;m-1,n} 
+ \beta_{s}^{(-)} \left(l-n+1\right)  s_{k,l;m,n-1}
\nonumber\\
&-&  \gamma_{s}^{(+)} m  s_{k,l;m,n}
- \gamma_{s}^{(-)} n  s_{k,l;m,n} 
%\nonumber\\
+ \gamma_{s}^{(+)} \left( m+1\right)  s_{k,l;m+1,n} 
+ \gamma_{s}^{(-)} \left( n+1\right)  s_{k,l;m,n+1},
\label{dskmdt}\\
\frac{d c_{k,l;m,n}}{dt} &=& - R_{k,l;m,n} c_{k,l;m,n}+ F_{k,l;m,n} s_{k,l;m,n} 
\nonumber\\
&-&  \beta_{c}^{(+)} \left(k-l-m\right)  c_{k,l;m,n}
- \beta_{c}^{(-)} \left(l-n\right)  c_{k,l;m,n} 
\nonumber\\
&+& \beta_{c}^{(+)} \left(k-l-m+1\right)  c_{k,l;m-1,n} 
+ \beta_{c}^{(-)} \left(l-n+1\right)  c_{k,l;m,n-1}
\nonumber\\
&-&  \gamma_{c}^{(+)} m  c_{k,l;m,n}
- \gamma_{c}^{(-)} n  c_{k,l;m,n} 
%\nonumber\\
+ \gamma_{c}^{(+)} \left( m+1\right)  c_{k,l;m+1,n} 
+ \gamma_{c}^{(-)} \left( n+1\right)  c_{k,l;m,n+1}.
\label{dikmdt}
\end{eqnarray}
In Eq.\ (\ref{dskmdt}), (\ref{dikmdt}), the first two terms account for the effect of a flip of a spin in a node with degree $k$ and $l$ attached negative links. In terms of Eq.\ \ref{rate_fxk}, the flip rate for a spin oriented down occupying a node with degree $k$,
attached $l$ negative links, with $m$ neighboring spins oriented up connected by positive links and $n$ neighboring spins oriented up connected by negative links is $F_{k,l;m,n}= f \left( m+l-n \left| k \right. \right)$ and that for an analogous spin oriented up is $R_{k,l;m,n}= f \left( k-l-m+n \left| k \right. \right)$. The remaining terms account for the average effect of the flips of spins in the neighboring nodes, irrespective of their degrees $k$ and numbers of attached negative links $l$. For example, the third term in Eq.\ (\ref{dskmdt}) accounts for the decrease of the concentration of spins oriented down in the above-mentioned nodes, with the above-mentioned numbers of neighboring spins oriented up, due to the flips of neighboring spins connected by positive links and oriented down (the number of such spins is $k-l-m$, and each flip of such spin changes $m$ to $m+1$, thus decreasing $s_{k,l;m,n}$ and increasing $s_{k,l;m+1,n}$), which occur at the average rate $\beta_{s}^{(+)}$. The rates $\beta_{s}^{(+)}, \ldots \gamma_{c}^{(-)}$ in Eq.\ (\ref{dskmdt}, \ref{dikmdt}) can be obtained as normalized double averages of the spin-flip rates $F_{k,l;m,n}$, $R_{k,l;m,n}$ over the degree distribution $P(k)$ and the binomial distribution $B_{k,l}(r)$ of the number $l$ of attached negative links. Their explicit form is given in Appendix A. It should be remembered that due to the constraints for the possible values of $m$, $n$ in Eq.\ (\ref{dskmdt}, \ref{dikmdt}) there is $s_{k,l;m-1,n} \equiv 0$ if $m-1<0, \ldots$ $c_{k,l;m,n+1}\equiv 0$ if $n+1>l$. The initial conditions are $s_{k,l;m,n}(0)= (1-c_0) B_{k-l,m}(c_0) B_{l,n} (c_0)$, $c_{k,l;m,n}(0)= c_0 B_{k-l,m}(c_0) B_{l,n} (c_0)$, where $0\le c_0 \le 1$ is arbitrary. 

The quantities of interest, which can be used to identify the possible phase transition from the PM to the FM state, e.g., the concentration $c$ of spins oriented up and the magnetization $m$, are evaluated as
\begin{equation}
    c(t) = \Big \langle \Big \langle \sum_{m=0}^{k-l}\sum_{n=0}^{l} c_{k,l;m,n}(t) \Big \rangle \Big \rangle,
    \label{c_total_AME}
\end{equation}
where the double average $\langle \langle \ldots \rangle\rangle$ is defined in Eq.\ (\ref{doubleav}), and $m=2c-1$. In the PM state, there is $c=1/2$, $m=0$, while in the two symmetric FM states there is $c<1/2$, $m<0$ or $c>1/2$, $m>0$.

The number of equations in the system (\ref{dskmdt}, \ref{dikmdt}) for each $k$ is $2 \sum_{l=0}^{k} (k-l+1)(l+1)= (k+1)(k+2)(k+3)/3$, which even for relatively small $k=10$ is $572$. The total number of equations (\ref{dskmdt}, \ref{dikmdt}) depends on the distribution $P(k)$, and in particular, for scale-free distributions with heavy tails increases dramatically. Thus, in this paper, only the $q$-voter model with independence on RRGs with $P(k)=\delta_{k, K}$ is investigated both using MC simulations and AMEs; the results for other, weakly heterogeneous networks with $\langle k\rangle =K$ are expected to be quantitatively similar.   

\subsection{Heterogeneous pair approximation}

\label{sec:theory_HePA}

The AMEs are a starting point for a certain formulation of the PA called AMEs-based heterogeneous PA \cite{Gleeson11, Gleeson13}, which was extended to the case of models with binary-state dynamics on weighted networks \cite{Peralta20, Unicomb18}. In this approximation, the following aggregate variables are considered, 
\begin{eqnarray}
    c_{k,l}(t)&=&\sum_{m=0}^{k-l}\sum_{n=0}^{l} c_{k,l;m,n}(t),\\
    \label{ckl}
    \vartheta_{k,l}^{(+)}&=&\frac{\sum_{m=0}^{k-l} \sum_{n=0}^{l}m s_{k,l;m,n}}{\sum_{m=0}^{k-l} \sum_{n=0}^{l}(k-l) s_{k,l;m;m}} = \frac{\sum_{m=0}^{k-l} \sum_{n=0}^{l}m s_{k,l;m,n}}{(k-l) \left( 1-c_{k,l}\right)},
    \label{thetaklp}\\
    \vartheta_{k,l}^{(-)}&=& \frac{\sum_{m=0}^{k-l} \sum_{n=0}^{l}n s_{k,l;m,n}}{l \left( 1-c_{k,l}\right)},
    \label{thetaklm}\\  
    \eta_{k,l}^{(+)}&=& \frac{\sum_{m=0}^{k-l} \sum_{n=0}^{l}m c_{k,l;m,n}}{(k-l)c_{k,l}},
    \label{etaklp}\\
    \eta_{k,l}^{(-)}&=& \frac{\sum_{m=0}^{k-l} \sum_{n=0}^{l}n c_{k,l;m,n}}{l c_{k,l}}.
    \label{etaklm}
\end{eqnarray}
The variable $c_{k,l}$ in Eq.\ (\ref{ckl}) represents the density of spins oriented up occupying nodes with degree $k$ and $l$ attached negative links. The variables $\vartheta_{k,l}^{(+)}$ in Eq.\ (\ref{thetaklp}) and $\vartheta_{k,l}^{(-)}$ in Eq.\ (\ref{thetaklm}) represent probabilities that a spin oriented down and occupying a node with degree $k$ and $l$ attached negative links has a neighboring spin oriented up and connected by a positive or negative link, respectively. The variables $\eta_{k,l}^{(+)}$ in Eq.\ (\ref{etaklp}) and $\eta_{k,l}^{(-)}$ in Eq.\ (\ref{etaklm}) represent analogous probabilities for a spin oriented up and occupying a node with degree $k$ and $l$ attached negative links.

The core of the AMEs-based heterogeneous PA is the assumption that for a spin oriented down and occupying a node with degree $k$ and $l$ attached negative links the numbers $m$, $n$ of its neighbors oriented up and connected by positive or negative links, respectively, obey binomial distributions $B_{k-l,m} \left(\vartheta^{(+)}_{k,l} \right)$, $B_{l,n}\left(\vartheta^{(-)}_{k,l}\right)$. Analogous distributions for a spin oriented up and occupying a node with degree $k$ and $l$ attached negative links are $B_{k-l,m} \left(\eta^{(+)}_{k,l} \right)$, $B_{l,n}\left(\eta^{(-)}_{k,l}\right)$. Then, the following approximations can be made,
\begin{eqnarray}
    s_{k,l;m,n} &\approx& \left( 1 -c_{k,l}\right)  B_{k-l,m}\left(\vartheta^{(+)}_{k,l} \right) B_{l,n}\left(\vartheta^{(-)}_{k,l}\right), 
    \label{sklmnPA}\\
    c_{k,l;m,n} &\approx& c_{k,l}  B_{k-l,m}\left(\eta^{(+)}_{k,l} \right) B_{l,n}\left(\eta^{(-)}_{k,l}\right).  
    \label{cklmnPA}
\end{eqnarray}
The above-mentioned approximations should also be made in the average rates $\beta_{s}^{(+)}, \ldots \gamma_{c}^{(-)}$, given by Eq.\ (\ref{betasp} - \ref{gammacm}) in Appendix A. This procedure, which is explained in Appendix B, yields approximate average rates $\bar{\beta}_{s}^{(+)}, \ldots \bar{\gamma}_{c}^{(-)}$, which are also given in Appendix B. 
Differentiating the definitions of $c_{k,l} \ldots \eta_{k,l}^{(-)}$, Eq.\ (\ref{dckldt} - \ref{detaklmdt}), with respect to time and using  Eq.\ (\ref{dskmdt}), (\ref{dikmdt}) with the approximations given by Eq.\ (\ref{sklmnPA}, \ref{cklmnPA}) yields the following 
system of equations for the time dependence of the aggregate variables in the AMEs-based heterogeneous PA,
\begin{eqnarray}
    \frac{dc_{k,l}}{dt} &=& 
    \sum_{m=0}^{k-l} \sum_{n=0}^{l}
    \left[
    -c_{k,l} R_{k,l;m,n} B_{k-l,m}\left(\eta^{(+)}_{k,l} \right) B_{l,n}\left(\eta^{(-)}_{k,l}\right) +
    \left( 1- c_{k,l}\right) F_{k,l;m,n} B_{k-l,m}\left(\vartheta^{(+)}_{k,l} \right) B_{l,n}\left(\vartheta^{(-)}_{k,l}\right) \right], \label{dckldt} \\
    \frac{d \vartheta^{(+)}_{k,l}}{dt} &=& 
    \sum_{m=0}^{k-l} \sum_{n=0}^{l}
    \left( \vartheta^{(+)}_{k,l} -\frac{m}{k-l}\right)
    \left[
    F_{k,l;m,n} B_{k-l,m}\left(\vartheta^{(+)}_{k,l} \right) B_{l,n}\left(\vartheta^{(-)}_{k,l}\right) -
   \frac{c_{k,l}}{1- c_{k,l}} R_{k,l;m,n} B_{k-l,m}\left(\eta^{(+)}_{k,l} \right) B_{l,n}\left(\eta^{(-)}_{k,l}\right) \right] \nonumber\\
   &+&\bar{\beta}_{s}^{(+)} \left(1- \vartheta^{(+)}_{k,l}\right) - \bar{\gamma}^{(+)}_{s} \vartheta^{(+)}_{k,l}, \label{dthetaklpdt}\\
    \frac{d \vartheta^{(-)}_{k,l}}{dt} &=& 
    \sum_{m=0}^{k-l} \sum_{n=0}^{l}
    \left( \vartheta^{(-)}_{k,l} -\frac{n}{l}\right)
    \left[
    F_{k,l;m,n} B_{k-l,m}\left(\vartheta^{(+)}_{k,l} \right) B_{l,n}\left(\vartheta^{(-)}_{k,l}\right) -
   \frac{c_{k,l}}{1- c_{k,l}} R_{k,l;m,n} B_{k-l,m}\left(\eta^{(+)}_{k,l} \right) B_{l,n}\left(\eta^{(-)}_{k,l}\right) \right] \nonumber\\
   &+&\bar{\beta}_{s}^{(-)} \left(1- \vartheta^{(-)}_{k,l}\right) - \bar{\gamma}^{(-)}_{s} \vartheta^{(-)}_{k,l},\label{dthetaklmdt}\\    
    \frac{d \eta^{(+)}_{k,l}}{dt} &=& 
    \sum_{m=0}^{k-l} \sum_{n=0}^{l}
    \left( \eta^{(+)}_{k,l} -\frac{m}{k-l}\right)
    \left[
    R_{k,l;m,n} B_{k-l,m}\left(\eta^{(+)}_{k,l} \right) B_{l,n}\left(\eta^{(-)}_{k,l}\right) -
   \frac{1- c_{k,l}}{c_{k,l}} F_{k,l;m,n} B_{k-l,m}\left(\vartheta^{(+)}_{k,l} \right) B_{l,n}\left(\vartheta^{(-)}_{k,l}\right) \right] \nonumber\\
   &+&\bar{\beta}_{c}^{(+)} \left(1- \eta^{(+)}_{k,l}\right) - \bar{\gamma}^{(+)}_{c} \eta^{(+)}_{k,l},\label{detaklpdt}\\
    \frac{d \eta^{(-)}_{k,l}}{dt} &=& 
    \sum_{m=0}^{k-l} \sum_{n=0}^{l}
    \left( \eta^{(-)}_{k,l} -\frac{n}{l}\right)
    \left[
    R_{k,l;m,n} B_{k-l,m}\left(\eta^{(+)}_{k,l} \right) B_{l,n}\left(\eta^{(-)}_{k,l}\right) -
   \frac{1- c_{k,l}}{c_{k,l}} F_{k,l;m,n} B_{k-l,m}\left(\vartheta^{(+)}_{k,l} \right) B_{l,n}\left(\vartheta^{(-)}_{k,l}\right) \right] \nonumber\\
   &+&\bar{\beta}_{c}^{(-)} \left(1- \eta^{(-)}_{k,l}\right) - \bar{\gamma}^{(-)}_{c} \eta^{(-)}_{k,l}, \label{detaklmdt}
\end{eqnarray}
Initial conditions are $c_{k,l}(0)=\vartheta_{k,l}^{(+)}(0)= \vartheta_{k,l}^{(-)}(0) = \eta_{k,l}^{(+)}(0) = \eta_{k,l}^{(-)}(0) =c(0)$, where $0\le c(0) \le 1$ is arbitrary. Taking into account Eq.\ (\ref{c_total_AME}), the total concentration of nodes occupied by spins oriented up $c$ is thus
\begin{equation}
    c(t) = \langle \langle  c_{k,l}(t) \rangle \rangle,
\end{equation}
and the magnetization is $m = 2c-1$. 

The AMEs-based heterogeneous PA is intermediate between the AMEs and the signed homogeneous PA previously used to study the $q$-voter model with independence on signed networks \cite{Krawiecki24}. It is less detailed than the AMEs due to the use of the aggregate variables (\ref{ckl} - \ref{etaklm}) and the binomial approximation (\ref{sklmnPA}, \ref{cklmnPA}). In turn, it is more detailed than the signed homogeneous PA since the aggregate variables $c_{k,l}, \ldots \eta_{k,l}^{(-)}$ depend on $k$, $l$ rather than being averages over the degree distribution $P(k)$ and the distribution of the number of attached negative links $B_{k,l}(r)$. Nevertheless, as shown in Sec.\ \ref{sec:results_MC_and_theory}, the AMEs-based heterogeneous PA yields predictions quantitatively indistinguishable from these of the signed homogeneous PA. 

%444444444444444444444444444444444444444444444444444444

\section{Results}

\label{sec:results}

\subsection{Details of Monte Carlo simulations}

\label{sec:results_MC_details}

In this section, results of MC simulations of the $q$-voter model with independence on signed networks are presented, and their results are compared with predictions of the AMEs and AMEs-based heterogeneous PA from Sec.\ \ref{sec:theory} concerning the FM transition. For reasons explained in Sec.\ \ref{sec:theory_AMEs}, only the model on RRGs with a relatively small degree of nodes $K$ was studied. 
Simulations were performed in a way described in more detail in Ref.\ \cite{Krawiecki24} on networks with the number of nodes $10^{3}\le N\le 10^{4}$ using a simulated annealing algorithm with random sequential updating of the agents' opinions, as described in Sec.\ \ref{sec:model}. Both PM and FM initial conditions were used, with random or uniform (up or down) initial distribution of spins, respectively. The results were averaged over $100-500$ (depending on $N$) realizations of the network and the distribution of the signs of links.

The order parameter for the FM transition is the absolute value of the magnetization
\begin{equation}
M=\left| \left[ \langle \frac{1}{N}\sum_{j=1}^{N} \sigma_{j}\rangle_{t}\right]_{av} \right|
 \equiv \left|\left[ \langle \tilde{m} \rangle_{t} \right]_{av} \right|,
\label{M}
\end{equation}
where $\tilde{m}$ denotes a momentary value of the magnetization at a given MCSS, $\langle \cdot \rangle_{t}$ denotes the time average for a model with a given realization of the network according to $P(k)=\delta_{k, K}$ and with a given associated distribution of the signs of links, and $\left[ \cdot \right]_{av}$ denotes average over different realizations of the network and different associated distributions of the signs of links. The order of the FM transition and the critical value of the independence parameter can be conveniently determined using the Binder cumulant $U^{(M)}$ vs.\ $p$ \cite{Binder97}, 
\begin{equation}
U^{(M)}=\frac{1}{2}\left[ 3-\frac{\langle \tilde{m}^{4} \rangle_{t}}{\langle \tilde{m}^{2}\rangle_{t}^{2}} \right]_{av},
\label{ULM}
\end{equation}
which has the property that for $p\rightarrow 0$ there is $U^{(M)}\rightarrow 1$ and for $p\rightarrow 1$ there is $U^{(M)}\rightarrow 0$. In the case of the second-order FM transition, this cumulant is a monotonically decreasing function of the independence parameter, while in the case of the first-order FM transition, it exhibits a negative minimum, which deepens and becomes sharper with an increasing number of nodes $N$. In the case of the second-order FM transition, the critical value of the independence parameter $p_{c, MC}^{(FM)}$ can be determined from the intersection point of the Binder cumulants for models with different numbers of agents $N$ \cite{Binder97}, simulated with PM or FM initial conditions. In the case of the first-order FM transition, it is sometimes possible to observe the hysteresis loop directly by measuring magnetization $M$ as a function of decreasing independence parameter for a model started in the PM phase to get $p_{c1, MC}^{(FM)}$, as well as as a function of increasing independence parameter for a model started in the FM phase to get $p_{c2, MC}^{(FM)}$; to obtain reliable results simulations of the model with the maximum number of nodes $N=10^4$ are utilized for this purpose. If the hysteresis loop is narrow, the critical value of the independence parameter, e.g., $p_{c1, MC}^{(FM)}$ again can be determined from the intersection point of the cumulants $U^{(M)}$ for models with different numbers of agents $N$, started in the PM phase. It should be noted that the critical values of the independence parameter $p_{c1,MC}^{(FM)}$, $p_{c2,MC}^{(FM)}$ or $p_{c,MC}^{(FM)}$ are expected to decrease fast with $q$ \cite{Jedrzejewski17, Krawiecki24}, thus the spin-flip rate (\ref{rate_fxk}) at the critical point also decreases. Hence, in practice, the FM transition in the model under study can be investigated using MC simulations only for small and moderate $q$ due to the prohibitively long simulation time needed to obtain reliable time averages for $M$ (\ref{M}), $U^{(M)}$ (\ref{ULM}).

\subsection{Numerical results and comparison with theoretical predictions}

\label{sec:results_MC_and_theory}

The $q$-voter model with independence on signed weakly heterogeneous networks was studied in Ref.\ \cite{Krawiecki24} for different parameters $\langle k\rangle$, $q$, including these mentioned below in this section. Results of MC simulations were compared with predictions of the MFA and homogeneous PA, in particular the signed homogeneous PA. In general, MC simulations reveal that for any $\langle k\rangle$ and $q\le \langle k\rangle \ge 2$, there is a maximum value $r_{\rm max}$ such that for fixed $r<r_{\rm max}$ the model shows the FM transition with decreasing $p$. For $q \le 5$, the transition is always second-order, and the critical value $p_{c,MC}^{(FM)}$ decreases with $r$. For $q \ge 6$ the transition is first-order for small $r$, with the lower and upper critical values $p_{c1,MC}^{(FM)}$, $p_{c2,MC}^{(FM)}$, respectively, decreasing and approaching each other with $r$, so the width of the hysteresis loop diminishes; for $p< p_{c1,MC}^{(FM)}$ only the FM phase is stable, for $p_{c1,MC}^{(FM)} < p< p_{c2,MC}^{(FM)}$ both PM and FM phases are stable and coexist, and for $p>p_{c2,MC}^{(FM)}$ only the PM phase is stable. Eventually, both critical lines on the $(r,p)$ phase plane can merge, and the hysteresis loop can disappear at the TCP at $r=r_{TCP,MC}$, and for $r>r_{TCP,MC}$, the transition becomes second-order, with the critical value $p_{c,MC}^{(FM)}$ again decreasing with $r$. For $\langle k \rangle \gg q$, the order of the FM transition, the critical value(s) of the independence parameter, and the location of the TCP are quantitatively correctly predicted by the signed homogeneous PA, and for $\langle k \rangle \rightarrow\infty$ also by the MFA \cite{Krawiecki24}. However, for $\langle k \rangle$ small and comparable with $q$ ($q\le \langle k \rangle$), predictions of the signed homogeneous PA become wrong: the critical value(s) of the independence parameter are overestimated for almost all $r$, and even the expected order of the FM transition is incorrect \cite{Krawiecki24}. In this section, the $q$-voter model with independence on signed random graphs with the parameters $\langle k \rangle$, $q$ in the latter range is investigated theoretically using more advanced and accurate approximations, the AMEs and the AMEs-based heterogeneous PA (Sec.\ \ref{sec:theory}). For reasons mentioned in Sec.\ \ref{sec:theory_AMEs}, only the model on signed RRGs with relatively small $K=10$ and different values of $q$ ($q =4, 6, 8$) is studied as an example. By comparison with the results of MC simulations, it is shown that predictions concerning the FM transition obtained using the AMEs are quantitatively correct in this case. In contrast, the AMEs-based heterogeneous PA yields predictions indistinguishable from those of the signed homogeneous PA, thus, in most cases, incorrect.

\begin{figure}
    \includegraphics[width=0.5\linewidth]{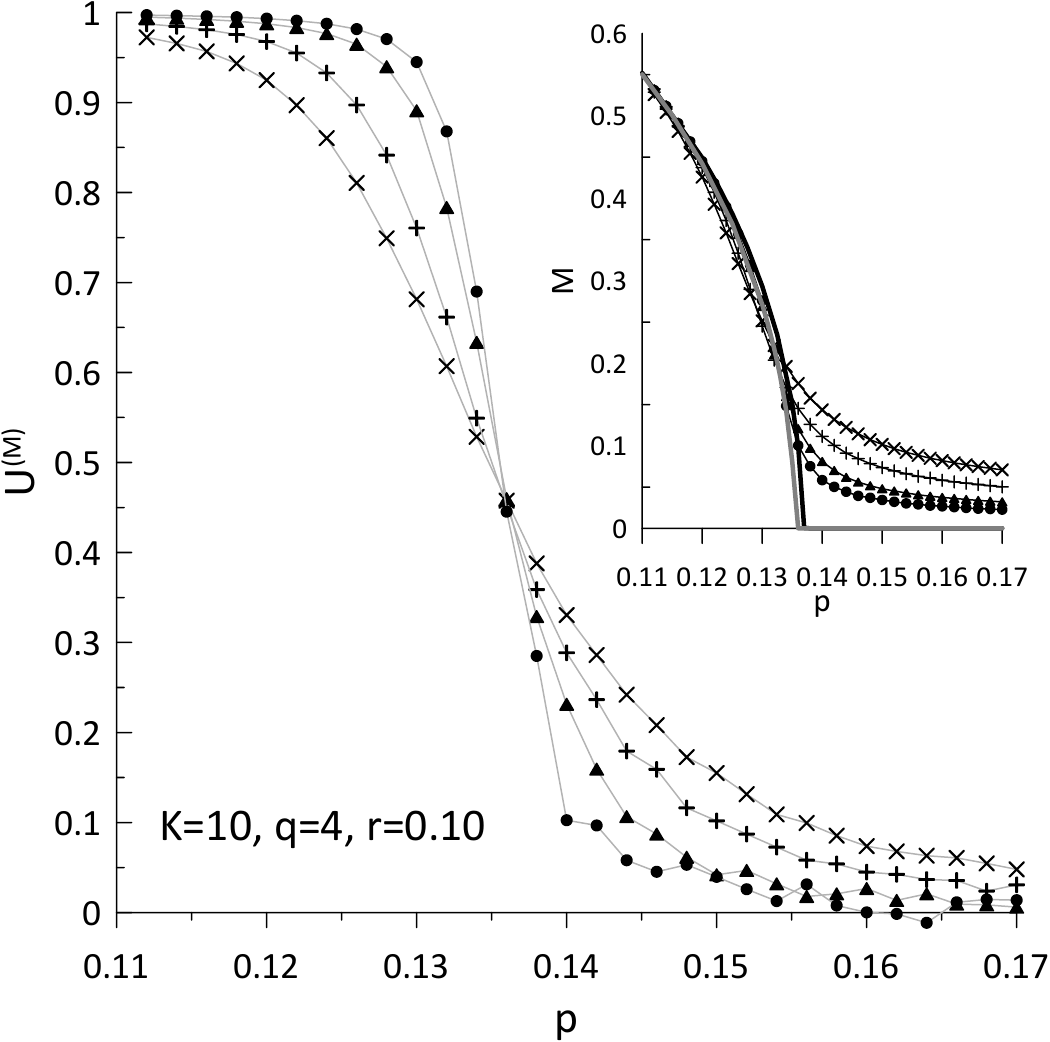}
    \caption{The Binder cumulants $U^{(M)}$ vs.\ $p$ from MC simulations of the $q$-voter model with independence on signed RRGs with $K=10$, $q=4$, $r=0.1$ for $N=10^3$ ($\times$), $N=2\cdot 10^3$ ($+$), $N=5\cdot 10^3$ ($\blacktriangle$),  $N=10^4$  ($\bullet$), gray solid lines are guides to the eyes. Inset: magnetization $M$ vs.\ $p$, symbols as above, thick black and gray lines show predictions of the AMEs and heterogeneous PA, respectively.}
    \label{fig:fig1}
\end{figure}

\begin{figure}
    \includegraphics[width=0.5\linewidth]{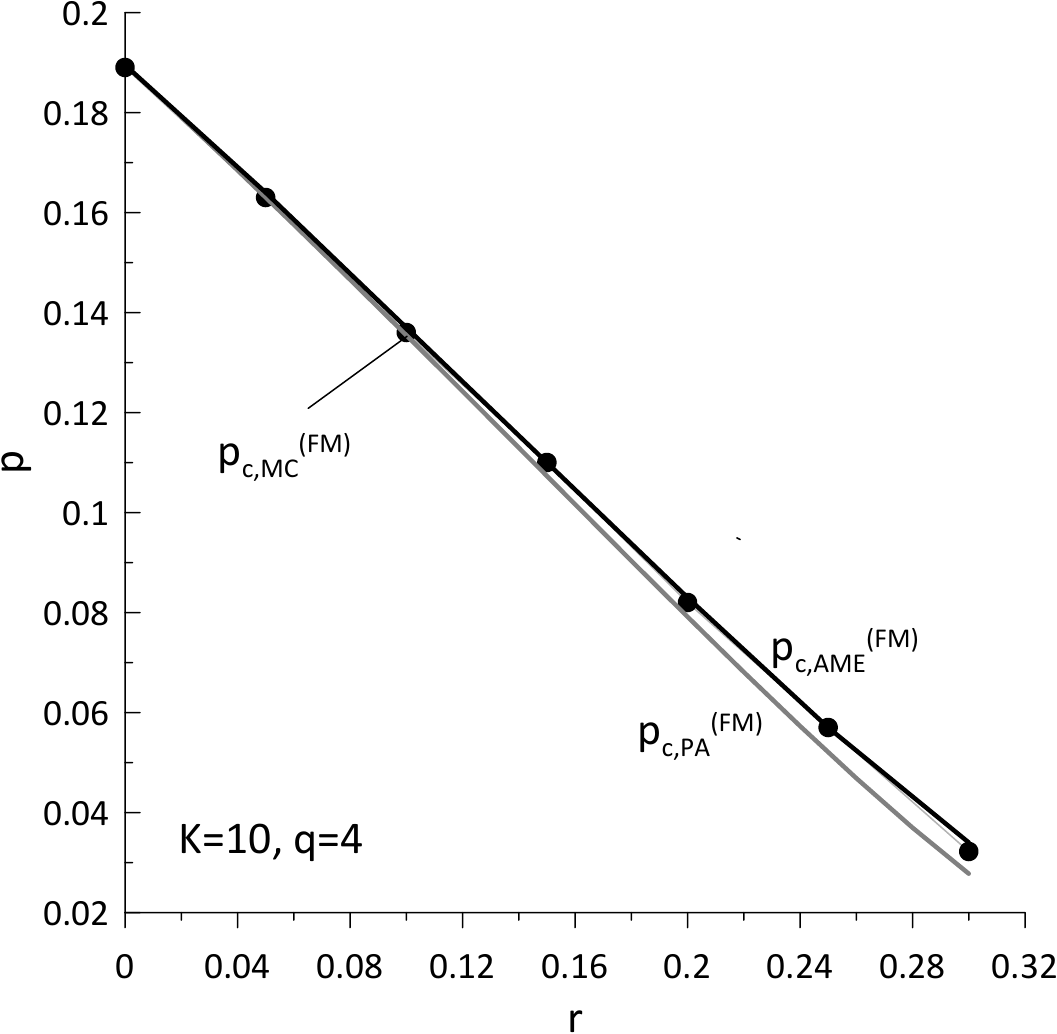} 
    \caption{Phase diagram for the $q$-voter model with independence on signed RRGs with $K=10$, $q=4$. The particular critical lines for the FM transition are labeled on the diagram. Symbols ($\bullet$) denote critical lines obtained from MC simulations for the second-order FM transition. Solid lines denote critical lines predicted by the AMEs (thick black line) and heterogeneous PA (thick gray line).}
    \label{fig:fig2}
\end{figure}

For $K=10$, $q=4$, and any $r$, the FM transition with decreasing $p$ observed in MC simulations of the model under study is second-order: for fixed $r$, the Binder cumulants $U^{(M)}$ decrease monotonically with $p$ and cross at the critical value of the independence parameter $p=p_{c,MC}^{(FM)}$ for different $N$ (Fig.\ \ref{fig:fig1}), and the latter value decreases almost linearly with $r$ (Fig.\ \ref{fig:fig2}). In this case, predictions of both AMEs and heterogeneous PA concerning the FM transition show good quantitative agreement with the results of MC simulations. The magnetization $M$ as a function of $p$ is quantitatively correctly predicted by both the AMEs and heterogeneous PA, although for large $r$, it is slightly underestimated by the latter approximation (Fig.\ \ref{fig:fig1}). Also for a whole range of $r$ where the FM transition occurs, the critical values of the independence parameter for the second-order transition $p_{c,AME}^{(FM)}$, $p_{c,PA}^{(FM)}$ predicted from the AMEs and heterogeneous PA, respectively, agree quantitatively with $p_{c,MC}^{(FM)}$ (Fig.\ \ref{fig:fig2}), although again for large $r$ the values of $p_{c,PA}^{(FM)}$ are slightly underestimated. 

\begin{figure}[h]
    \includegraphics[width=0.5\linewidth]{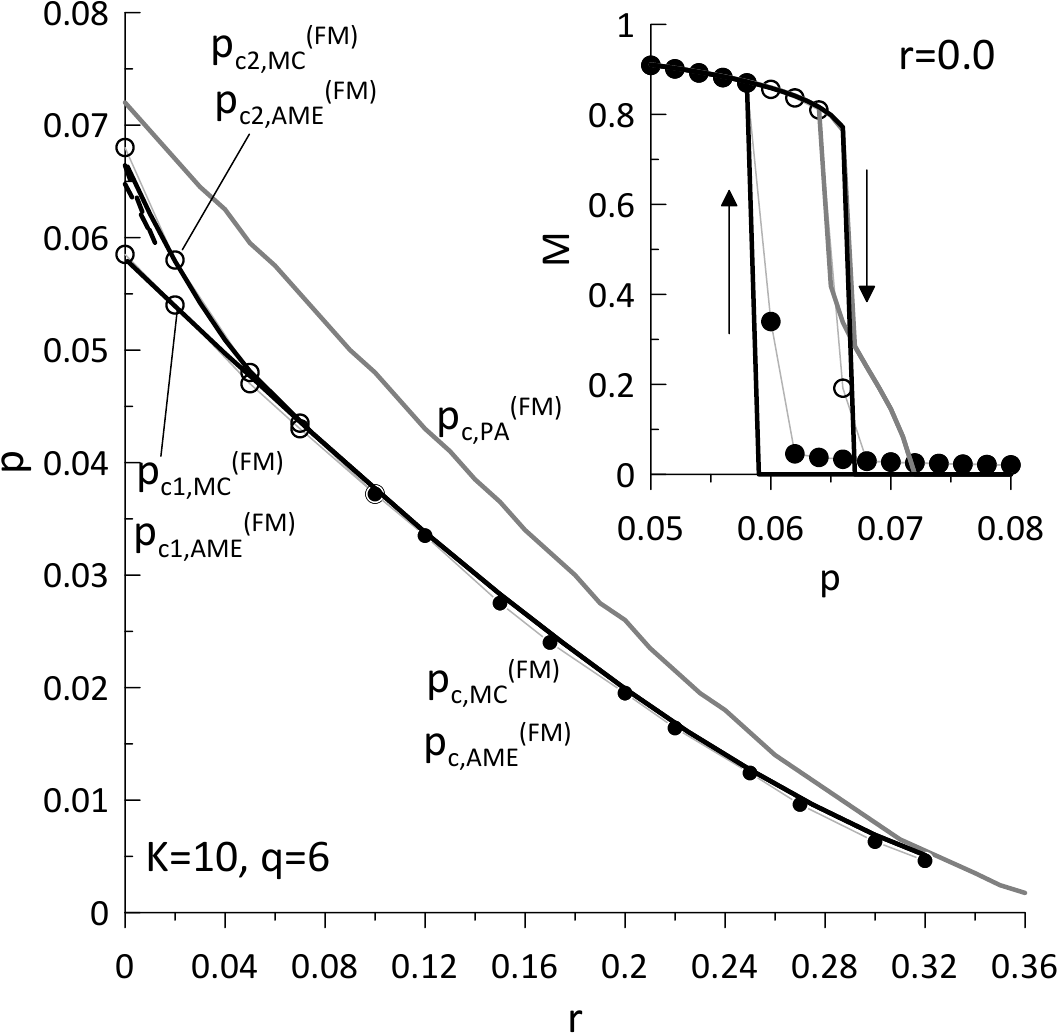}
    \caption{As in Fig.\ \ref{fig:fig2}, but for $K=10$, $q=6$. Symbols denote critical lines obtained from MC simulations for the first-order ($\circ$) and second-order ($\bullet$) FM transition. Thin gray lines are guides to the eye. Solid lines denote critical lines predicted by the AMEs (thick black line) and heterogeneous PA (thick gray line). Dashed lines denote the lower and upper critical lines for the additional discontinuous FM transition predicted by the heterogeneous PA, occurring for $p<p_{c,PA}^{(FM)}$, i.e., below the continuous FM transition, as shown in the inset.  Inset: magnetization $M$ vs.\ $p$ from MC simulations of the model with $r=0.0$ started with PM conditions and decreasing $p$ ($\bullet$) as well as with FM conditions and increasing $p$ ($\circ$),  thick black and gray lines show predictions of the AMEs and heterogeneous PA, respectively (cf.\ Fig.\ 7 in Ref.\ \cite{Krawiecki24}).}
    \label{fig:fig3}
\end{figure}

\begin{figure}[h]
    \includegraphics[width=0.5\linewidth]{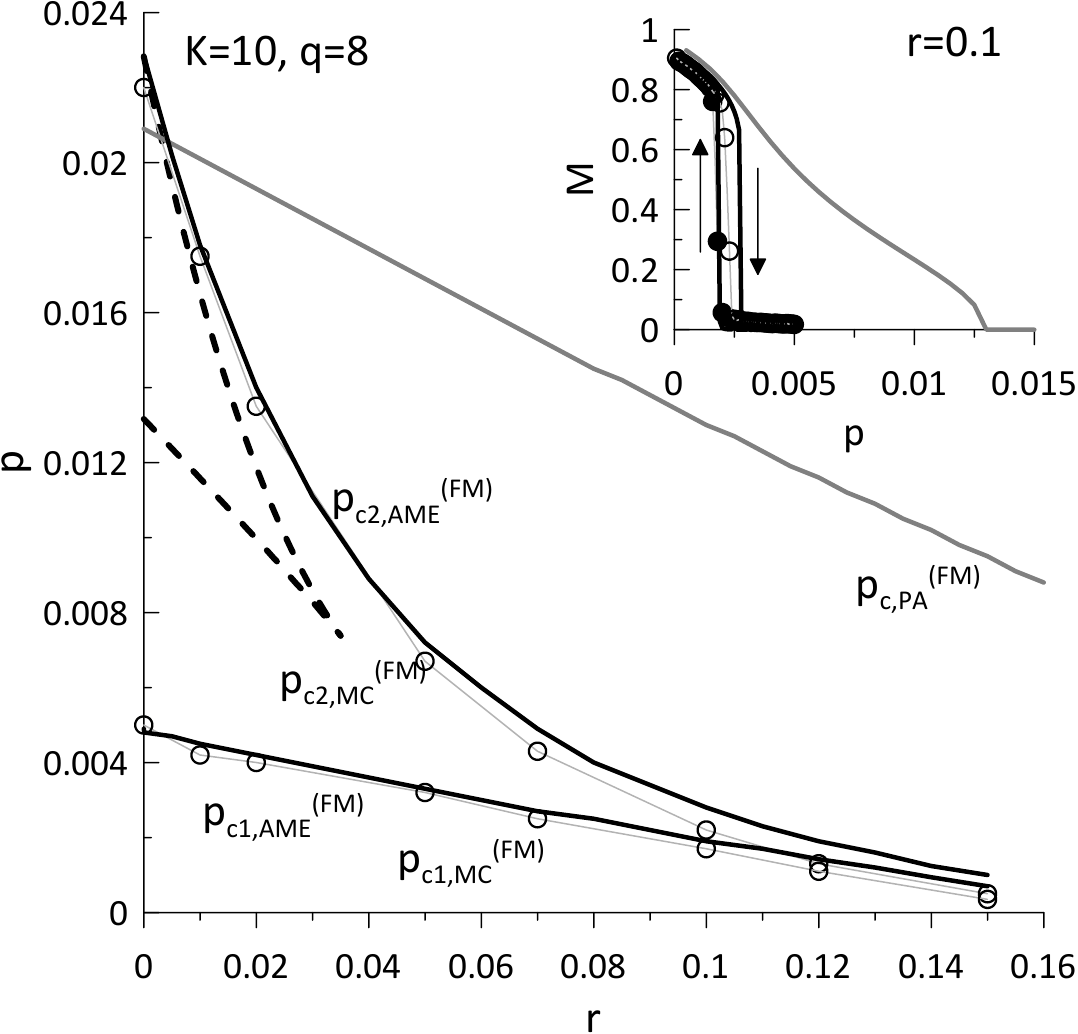}
    \caption{As in Fig.\ \ref{fig:fig3}, but for $K=10$, $q=8$. Inset: as in Fig.\ \ref{fig:fig3},  for the model with $r=0.1$ (cf.\ Fig.\ 8 in Ref.\ \cite{Krawiecki24}).}
    \label{fig:fig4}
\end{figure}

For both $q=6$ and $q=8$, MC simulations of the model under study reveal the first-order FM transition with the hysteresis loop for small $r$ (Fig.\ \ref{fig:fig3}, \ref{fig:fig4}). For larger $r$, the width of the hysteresis loop decreases, and for $q=6$, the transition eventually becomes second-order (Fig.\ \ref{fig:fig3}), while for $q=8$ it remains first-order (Fig.\ \ref{fig:fig4}). For $q=6$, the lower and upper critical values $p_{c1,MC}^{(FM)}$, $p_{c2,MC}^{(FM)}$ of the independence parameter for the first-order FM transition, obtained from MC simulations, approach each other with increasing $r$ and eventually merge at a TCP, where the width of the hysteresis loop becomes zero. For $r$ just above this TCP, in MC simulations, the FM transition still exhibits some features of the first-order transition: although the hysteresis loop is not observed directly, the Binder cumulants $U^{(M)}$ for different $N$ cross at one point corresponding to the critical value $p_{c,MC}^{(FM)}$ and exhibit negative minima as functions of $p$ which become deeper with an increasing number of nodes \cite{Krawiecki24}. Only for still higher values of $r$ does the FM transition observed in MC simulations become continuous. 

For $q=6$, and in the whole range of $r$, the order of the FM transition in the model under study predicted by the AMEs, the width of the hysteresis loop and the lower and upper critical values $p_{c1,AME}^{(FM)}$, $p_{c2,AME}^{(FM)}$ of the independence parameter for the first-order FM transition, the critical value $p_{c,AME}^{(FM)}$ for the second-order FM transition, as well as the location of the TCP separating the critical curves for the FM transitions of different orders on the $(r,p)$ plane in general exhibit good quantitative agreement with these obtained from the MC simulations (Fig.\ \ref{fig:fig3}). Similar agreement occurs for $q=8$, where only for larger $r$ small differences appear between the dependence of the magnetization on $p$ as well as the width and location of the hysteresis loop observed in the MC simulations and predicted by the AMEs (Fig.\ \ref{fig:fig4}). In contrast, both for $q=6$ and $q=8$, predictions of the heterogeneous PA are even qualitatively incorrect: in the whole range of $r$, this approximation predicts only the occurrence of the second-order FM transition with decreasing $p$, and the critical values $p_{c,PA}^{(FM)}$ of the independence parameter are significantly overestimated (Fig.\ \ref{fig:fig3}, \ref{fig:fig4}). Moreover, according to the heterogeneous PA at $p< p_{c,PA}^{(FM)}$ another discontinuous phase transition can appear between two FM phases with low and high magnetization (Fig.\ \ref{fig:fig3}, \ref{fig:fig4}), which is not observed in MC simulations. The range of parameters $r$, $p$ for the occurrence of this additional FM transition is much narrower than that for the first-order FM transition observed in MC simulations. 

It may be verified that for the model under study the above-mentioned predictions of the AMEs-based heterogeneous PA are indistinguishable from those of the signed heterogeneous PA \cite{Krawiecki24}; in particular, in the cases with $K=10$, $q=6$ and $q=8$ this can be seen by comparing Fig.\ \ref{fig:fig3} and \ref{fig:fig4} with the corresponding Fig.\ 7 and Fig.\ 8 in Ref.\ \cite{Krawiecki24}. It should be mentioned that in a related study of the $q$-neighbor Ising model on multiplex networks with partial overlap of nodes, predictions of the appropriate homogeneous PA and AMEs-based heterogeneous PA are also identical \cite{Krawiecki23}. Thus, for the $q$-voter model with independence on signed networks with $\langle k\rangle$ comparable with $q$, the AMEs-based heterogeneous PA is not accurate enough to quantitatively predict the FM transition's properties.

%555555555555555555555555555555555555555555555555555

\section{Summary and conclusions}

\label{sec:summary}

In this paper, it was shown that the AMEs, modified to take into account the presence of both reinforcing (positive) and antagonistic (negative) interactions, quantitatively correctly predict the properties of the FM transition in the $q$-voter model with independence on signed networks observed in MC simulations. In particular, this is true for the model on networks with a small mean degree of nodes $\langle k \rangle$ comparable with the size of the $q$-neighborhood, where other approximations, e.g., the signed homogeneous PA, yield predictions quantitatively or even qualitatively incorrect. Thus, the use of the AMEs solves the problem of obtaining reliable theoretical predictions concerning the FM transition in the latter case. In contrast, predictions of the AMEs-based heterogeneous PA, derived from the AMEs by using aggregate variables, for the model under study are indistinguishable from those of the signed homogeneous PA, thus again incorrect in the case of the model on networks with $\langle k \rangle$ small and comparable with $q$. This does not preclude a possibility that predictions of more advanced forms of heterogeneous PA \cite{Pugliese09}, extended to the case of models on signed networks, can better reproduce the properties of the FM transition in the model under study observed in MC simulations, but this issue is beyond the scope of this paper. The form of the AMEs (\ref{dskmdt}, \ref{dikmdt}) is general, so they can be applied to other models with binary-state dynamics on signed networks characterized by different spin-flip rates. In particular, they can be applied to related models for social opinion formation, e.g., the majority vote or the $q$-neighbor Ising model, to improve theoretical analysis of the possible FM transition. This problem is left for future research.  

%AAAAAAAAAAAAAAAAAAAAAAAAAAAAAAAAAAAAAAAAAAAAAAAAAAA

\section*{Appendix A}

Denoting the double average over the distributions $P(k)$ and $B_{k,l}(r)$ as  
\begin{equation}
    \langle \langle \ldots \rangle \rangle \equiv \sum_{k}P(k) \sum_{l=0}^{k} B_{k,l}(r) \ldots,
    \label{doubleav}
\end{equation}
the average rates in Eq.\ (\ref{dskmdt},\ref{dikmdt}) are
\begin{eqnarray}
    \beta_{s}^{(+)} &=& \frac{\bigl \langle \bigl\langle \sum_{m=0}^{k-l}\sum_{n=0}^{l} (k-l-m) F_{k,l;m,n} s_{k,l;m,n} \bigr\rangle \bigr\rangle}{\bigl\langle \bigl\langle \sum_{m=0}^{k-l}\sum_{n=0}^{l} (k-l-m) s_{k,l;m,n}\bigr\rangle \bigr\rangle}, \label{betasp}\\
    \beta_{s}^{(-)} &=&  \frac{\big\langle \big\langle \sum_{m=0}^{k-l}\sum_{n=0}^{l} (l-n) F_{k,l;m,n} s_{k,l;m,n}\big\rangle \big\rangle}{\big\langle \big\langle \sum_{m=0}^{k-l}\sum_{n=0}^{l} (l-n) s_{k,l;m,n}\big\rangle \big\rangle}, \label{betasm}\\
    \gamma_{s}^{(+)} &=& \frac{\big\langle \big\langle \sum_{m=0}^{k-l}\sum_{n=0}^{l} (k-l-m) R_{k,l;m,n} c_{k,l;m,n}\big\rangle \big\rangle}{\big\langle \big\langle \sum_{m=0}^{k-l}\sum_{n=0}^{l} (k-l-m) c_{k,l;m,n}\big\rangle \big\rangle},\label{gammasp}\\ \gamma_{s}^{(-)} &=& \frac{\big\langle \big\langle \sum_{m=0}^{k-l}\sum_{n=0}^{l} (l-n) R_{k,l;m,n} c_{k,l;m,n}\big\rangle \big\rangle}{\big\langle \big\langle \sum_{m=0}^{k-l}\sum_{n=0}^{l} (l-n) c_{k,l;m,n}\big\rangle \big\rangle}, \label{gammasm} \\
    \beta_{c}^{(+)} &=& \frac{\big\langle\big\langle \sum_{m=0}^{k-l}\sum_{n=0}^{l} m F_{k,l;m,n} s_{k,l;m,n}\big\rangle \big\rangle}{\big\langle \big\langle \sum_{m=0}^{k-l}\sum_{n=0}^{l} m s_{k,l;m,n}\big\rangle \big\rangle},\label{betacp}\\
    \beta_{c}^{(-)} &=& \frac{\big\langle \big\langle \sum_{m=0}^{k-l}\sum_{n=0}^{l} n F_{k,l;m,n} s_{k,l;m,n}\big\rangle \big\rangle}{\big\langle \big\langle \sum_{m=0}^{k-l}\sum_{n=0}^{l} n s_{k,l;m,n}\big\rangle \big\rangle},\label{betacm}\\
    \gamma_{c}^{(+)} &=& \frac{\big\langle \big\langle \sum_{m=0}^{k-l}\sum_{n=0}^{l} m R_{k,l;m,n} c_{k,l;m,n}\big\rangle \big\rangle}{\big\langle \big\langle \sum_{m=0}^{k-l}\sum_{n=0}^{l} m c_{k,l;m,n}\big\rangle \big\rangle}, \label{gammacp}\\    
    \gamma_{c}^{(-)} &=& \frac{\big\langle \big\langle \sum_{m=0}^{k-l}\sum_{n=0}^{l} n R_{k,l;m,n} c_{k,l;m,n}\big\rangle \big\rangle}{\big\langle \big\langle \sum_{m=0}^{k-l}\sum_{n=0}^{l} n c_{k,l;m,n}\big\rangle \big\rangle}.  \label{gammacm} 
\end{eqnarray}
For RRGs $P(k)=\delta_{k,K}$, thus there is no summation over $k$ in Eq.\ (\ref{doubleav}). 

%BBBBBBBBBBBBBBBBBBBBBBBBBBBBBBBBBBBBBBBBBBBBBBBBBBBB

\section*{Appendix B}

Making the substitutions given by Eq.\ (\ref{sklmnPA}, \ref{cklmnPA}) in the definitions of the average rates $\beta_{s}^{(+)}, \ldots \gamma_{c}^{(-)}$ given by Eq.\ (\ref{betasp} - \ref{gammacm}) yields approximate rates used in the AMEs-based heterogeneous PA in Eq.\ (\ref{dckldt} - \ref{detaklmdt}), denoted as $\bar{\beta}_{s}^{(+)}, \ldots \bar{\gamma}_{c}^{(-)}$. For example, the rate $\beta_{s}^{(+)}$ in Eq.\ (\ref{betasp}) is approximated as
\begin{eqnarray}
    \bar{\beta}_{s}^{(+)} &=& \frac{\Big\langle \Big\langle  \left( 1 -c_{k,l}\right) \sum_{m=0}^{k-l}\sum_{n=0}^{l}  B_{k-l,m}\left(\vartheta^{(+)}_{k,l} \right) B_{l,n}\left(\vartheta^{(-)}_{k,l}\right) (k-l-m) F_{k,l;m,n}\Big\rangle\Big\rangle}{\Big\langle \Big\langle  \left( 1 -c_{k,l}\right) \sum_{m=0}^{k-l}\sum_{n=0}^{l}  B_{k-l,m}\left(\vartheta^{(+)}_{k,l} \right) B_{l,n}\left(\vartheta^{(-)}_{k,l}\right) (k-l-m)\Big\rangle\Big\rangle} \nonumber\\
    &=& \frac{\Big\langle \Big\langle  \left( 1 -c_{k,l}\right) \sum_{m=0}^{k-l}\sum_{n=0}^{l}  B_{k-l,m}\left(\vartheta^{(+)}_{k,l} \right) B_{l,n}\left(\vartheta^{(-)}_{k,l}\right) (k-l-m) F_{k,l;m,n}\Big\rangle\Big\rangle}{\Big\langle \Big\langle  \left( 1 -c_{k,l}\right) (k-l)\left( 1-\vartheta_{k,l}^{(+)}\right)\Big\rangle\Big\rangle}, 
    \label{barbetasp}
\end{eqnarray}
etc. 

Using the identities for binomial coefficients ${m+n \choose q}= \sum_{z=0}^{q}{n \choose z}{m \choose q-z}$, ${k-l \choose m+q-z}{m+q-z \choose q-z} = {k-l \choose q-z}{k-l-q+z\choose m}$ and ${k \choose l}{k-l \choose q-z}{l \choose z}= {k \choose q}{q \choose z}{k-q \choose l-z}$ summations in Eq.\ (\ref{barbetasp}) as well as in analogous equations defining the remaining rates $\bar{\beta}_{s}^{(-)}\ldots \bar{\gamma}_{c}^{(-)}$ can be partly performed. Introducing the following notation for the triple averages,
\begin{eqnarray}
    \langle \langle \langle \ldots \rangle \rangle \rangle_{\beta} &\equiv& \sum_{k}P(k)\sum_{z=0}^{q}{q\choose z}\sum_{l=0}^{k-q}B_{k-q,l}(r) \left[r\left( 1- \vartheta_{k,l+z}^{(-)}\right) \right]^z \left[(1-r) \vartheta_{k,l+z}^{(+)}\right]^{q-z} \ldots \label{tripleavbeta}\\
    \langle \langle \langle \ldots \rangle \rangle \rangle_{\gamma} &\equiv& \sum_{k}P(k)\sum_{z=0}^{q}{q\choose z}\sum_{l=0}^{k-q}B_{k-q,l}(r) \left[r \eta_{k,l+z}^{(-)} \right]^z \left[(1-r) \left( 1- \eta_{k,l+z}^{(+)}\right)\right]^{q-z} \ldots, \label{tripleavgamma}       
\end{eqnarray}
it is obtained that
\begin{eqnarray}
     \bar{\beta}_{s}^{(+)} &=&  \frac{ (1-p)\Bigl \langle \Bigl \langle \Bigl \langle \left( 1 -c_{k,l+z} \right) (k-q-l) \left( 1- \vartheta_{k,l+z}^{(+)}\right) \Bigr \rangle \Bigr\rangle \Bigr\rangle_{\beta} + (p/2) \Bigl\langle \Bigl\langle  \left( 1 -c_{k,l}\right) (k-l) \left( 1-\vartheta_{k,l}^{(+)} \right) \Bigr\rangle \Bigr\rangle }{\Bigl\langle \Bigl\langle  \left( 1 -c_{k,l}\right) (k-l)\left( 1-\vartheta_{k,l}^{(+)}\right) \Bigr\rangle \Bigr\rangle}, \\
     \bar{\beta}_{s}^{(-)} &=&  \frac{ (1-p)\Big\langle \Big\langle \Big\langle \left( 1 -c_{k,l+z} \right) \left[ l \left( 1- \vartheta_{k,l+z}^{(-)}\right) +z \right]\Big\rangle\Big\rangle \Big\rangle_{\beta} + (p/2) \Big\langle\Big\langle  \left( 1 -c_{k,l}\right) l \left( 1-\vartheta_{k,l}^{(-)} \right) \Big\rangle \Big\rangle }{\Big\langle \Big\langle  \left( 1 -c_{k,l}\right) l\left( 1-\vartheta_{k,l}^{(-)}\right) \Big\rangle\Big\rangle}, \\
     \bar{\gamma}_{s}^{(+)} &=&  \frac{ (1-p)\Big\langle \Big\langle \Big\langle c_{k,l+z} \left[ (k-q-l) \left( 1- \eta_{k,l+z}^{(+)}\right) +q-z \right]\Big\rangle\Big\rangle \Big\rangle_{\gamma} + (p/2) \Big\langle\Big\langle c_{k,l} (k-l) \left( 1-\eta_{k,l}^{(+)} \right) \Big\rangle \Big\rangle }{\Big\langle \Big\langle  c_{k,l} (k-l)\left( 1-\eta_{k,l}^{(+)}\right) \Big\rangle\Big\rangle}, \\
     \bar{\gamma}_{s}^{(-)} &=&  \frac{ (1-p)\Big\langle \Big\langle \Big\langle c_{k,l+z} l \left( 1- \eta_{k,l+z}^{(-)}\right) \Big\rangle\Big\rangle \Big\rangle_{\gamma} + (p/2) \Big\langle\Big\langle  c_{k,l} l \left( 1-\eta_{k,l}^{(-)} \right) \Big\rangle \Big\rangle }{\Big\langle \Big\langle  c_{k,l} l\left( 1-\eta_{k,l}^{(-)}\right) \Big\rangle\Big\rangle}, \\
     \bar{\beta}_{c}^{(+)} &=&  \frac{ (1-p)\Big\langle \Big\langle \Big\langle \left( 1 -c_{k,l+z} \right) \left[ (k-q-l)  \vartheta_{k,l+z}^{(+)} +q-z\right]\Big\rangle\Big\rangle \Big\rangle_{\beta} + (p/2) \Big\langle\Big\langle  \left( 1 -c_{k,l}\right) (k-l) \vartheta_{k,l}^{(+)} \Big\rangle \Big\rangle }{\Big\langle \Big\langle  \left( 1 -c_{k,l}\right) (k-l)\vartheta_{k,l}^{(+)} \Big\rangle\Big\rangle}, \\ 
     \bar{\beta}_{c}^{(-)} &=&  \frac{ (1-p)\Big\langle \Big\langle \Big\langle \left( 1 -c_{k,l+z} \right) l \vartheta_{k,l+z}^{(-)} \Big\rangle\Big\rangle \Big\rangle_{\beta} + (p/2) \Big\langle\Big\langle  \left( 1 -c_{k,l}\right) l \vartheta_{k,l}^{(-)} \Big\rangle \Big\rangle }{\Big\langle \Big\langle  \left( 1 -c_{k,l}\right) l\vartheta_{k,l}^{(-)} \Big\rangle\Big\rangle},\\
     \bar{\gamma}_{c}^{(+)} &=&  \frac{ (1-p)\Big\langle \Big\langle \Big\langle c_{k,l+z}  (k-q-l) \eta_{k,l+z}^{(+)} \Big\rangle\Big\rangle \Big\rangle_{\gamma} + (p/2) \Big\langle\Big\langle c_{k,l} (k-l) \eta_{k,l}^{(+)} \Big\rangle \Big\rangle }{\Big\langle \Big\langle  c_{k,l} (k-l)\eta_{k,l}^{(+)} \Big\rangle\Big\rangle}, \\
     \bar{\gamma}_{c}^{(-)} &=&  \frac{ (1-p)\Big\langle \Big\langle \Big\langle c_{k,l+z} \left[ l \eta_{k,l+z}^{(-)}+z\right] \Big\rangle\Big\rangle \Big\rangle_{\gamma} + (p/2) \Big\langle\Big\langle c_{k,l} l \eta_{k,l}^{(-)} \Big\rangle \Big\rangle }{\Big\langle \Big\langle  c_{k,l} l \eta_{k,l}^{(-)} \Big\rangle\Big\rangle},
\end{eqnarray}
where, again, the double average $\langle\langle \ldots\rangle\rangle$ is given by Eq.\ (\ref{doubleav}). For RRGs $P(k)=\delta_{0,K}$, thus there is no summation over $k$ in Eq.\ (\ref{doubleav}, \ref{tripleavbeta}, 
 \ref{tripleavgamma}). 

%%%%%%%%%%%%%%%%%%%%%%%%%%%%%%%%%%%%%%%%%%%%%%%%%%%%%%%%%
%%%%%%%%%%%%%%%%%%%%%%%%%%%%%%%%%%%%%%%%%%%%%%%%%%%%%%%%%

\end{document}